\documentclass[aps,prl,twocolumn,amsmath,amssymb,superscriptaddress,nofootinbib]{revtex4-2}

\usepackage[utf8]{inputenc}
\usepackage{mathtools,bm}
\usepackage{microtype}
\usepackage{xcolor}
\usepackage{hyperref}
\hypersetup{colorlinks=true, linkcolor=black, citecolor=black, urlcolor=black}
\usepackage{nameref} 
\allowdisplaybreaks[1] 

\newcounter{secpara}
\renewcommand\thesecpara{\arabic{secpara}}

\newcommand{\PT}{\Pi_T^{(D)}}
\newcommand{\AdV}{\mathrm{Ad}_{V}} 

\makeatletter
\newcommand{\psec}[2]{%
  \refstepcounter{secpara}
  \phantomsection
  \def\@currentlabelname{#1}
  \paragraph*{\thesecpara.\;#1}\label{#2}%
}
\makeatother

\newcommand{\Sref}[1]{\textit{\nameref*{#1}}}
\newcommand{\Snameref}[1]{\Sref{#1}} 

\newcommand{\tr}{\mathrm{tr}}
\newcommand{\ii}{\mathrm{i}}
\newcommand{\dd}{\mathrm{d}}
\newcommand{\RR}{\mathbb{R}}
\newcommand{\ZZ}{\mathbb{Z}}
\newcommand{\cD}{\mathcal{D}}
\newcommand{\cJ}{\mathcal{J}}
\newcommand{\norm}[1]{\left\lVert #1\right\rVert}
\newcommand{\ip}[2]{\left\langle #1,#2\right\rangle}

\usepackage{orcidlink}
\usepackage{graphicx}
\usepackage{tikz}
\usetikzlibrary{arrows.meta,calc}

\begin{document}

\title{Mass Without Mass from a Berry--Shifted SU(3) Holonomy Rotor}

\author{Ahmed Farag Ali\orcidlink{0000-000X-XXXX-XXXX}}
\affiliation{Essex County College, 303 University Ave, Newark, NJ 07102, United States}
\affiliation{Department of Physics, Benha University, Benha 13518, Egypt}
\email{aali29@essex.edu}

\begin{abstract}
We identify a local, gauge-invariant mechanism that generates a finite spectral scale in pure SU(3) Yang--Mills theory on a punctured three-ball. Fixing a $\mathbb{Z}_3$ center sector isolates a single gauge-invariant holonomy angle whose Berry shift produces a quantum rotor with strictly nonzero level spacing. Gauss law is enforced by a covariant Dirichlet Helmholtz projector built from the Dirichlet inverse of the covariant scalar Laplacian with relative boundary conditions. The slow holonomy mode is chosen variationally as the minimizer of transverse electric energy under the holonomy constraint, yielding an inertia \emph{independent of the gauge representative} with linear domain-size scaling and a controlled commutator-dominated regime. We prove projector stability and derive an adiabatic variational upper bound on the first positive Yang--Mills eigenvalue, with error controlled by the transverse vector gap of the covariant Laplacian on divergence-free one-forms. A femtometer-scale benchmark at realistic coupling gives an upper bound at a hadronic ($\sim 1\,$GeV) scale.
In Wilczek's sense this realizes ``mass without mass'': no explicit mass term or Higgs field is introduced, and the nonzero level spacing is fixed by gauge invariance, topology, and the chosen center sector. The present results are derived on a finite domain; interpreting the length $R$ in Minkowski space requires an additional physical input (e.g.\ as a local confinement length), which we make explicit.
\end{abstract}

\maketitle

\psec{Introduction and context}{sec:intro}
“Mass without mass’’~\cite{Wilczek1999,Wilczek2000} highlights the emergence of spectral scales from gauge dynamics and topology alone. Confinement is the canonical example~\cite{Wilson1974,thooft1978,Greensite2011}, neatly organized by generalized (one-form/center) symmetries~\cite{Gaiotto2015,Kapustin2006}. Related SU(3)-based programs have explored confinement volumes as inputs to the cosmological constant and constants of nature~\cite{AliEPL2025,AliSymmetry2025}. Here we present a local, mass-parameter-free mechanism in pure SU(3) Yang--Mills on a punctured three-ball: fixing a $\mathbb{Z}_3$ center sector isolates a single gauge-invariant holonomy angle whose $\mathbb{Z}_3$ Berry shift produces a quantum rotor with strictly nonzero level spacing. The only dimensional input is the infrared completion length $R$ of the domain; the nontrivial content is that, once a center sector is fixed, the $\ZZ_3$ Berry shift and Gauss law force a nonzero rotor spacing with a computable dimensionless coefficient. Gauss law is enforced by a covariant Dirichlet Helmholtz projector built from the Dirichlet inverse of the covariant scalar Laplacian with relative boundary conditions, and the slow holonomy mode is selected variationally as the minimizer of transverse electric energy under the holonomy constraint. This yields a canonical, gauge-representative\,--\,independent inertia with linear domain-size scaling and a controlled commutator-dominated regime. We establish projector stability and an adiabatic variational upper bound on the first positive Yang--Mills eigenvalue, with error governed by the transverse vector gap. Throughout we use standard non-Abelian Yang--Mills conventions~\cite{YM}: $\tr(T^aT^b)=\tfrac12\delta^{ab}$, Minkowski signature $(-,+,+,+)$ (so that $\tr(F_{0i}F_{0i})\ge0$), and the Wilson-line definition $U_\gamma[A]=\mathcal P\exp(\ii g\!\oint_\gamma A)$ with $g^2=4\pi\alpha_s(\mu)$. We also choose $\cJ$ in the SU(3) \emph{coweight lattice} so that $e^{2\pi \ii \, \cJ}\in Z(\mathrm{SU(3)})\cong\ZZ_3$, implying a $2\pi$ periodicity \emph{up to a center element}; in a fixed $\ZZ_3$ sector this yields the twist~\eqref{eq:twist}. We write $\AdV X := VXV^{-1}$.

Main claim and scope (finite-domain ``mass without mass''). 
On the compact punctured domain $\cD=B_R\setminus\mathrm{Tub}(\Gamma)$, fixing a $\ZZ_3$ center sector isolates a gauge-invariant collective coordinate (the holonomy angle $\alpha$) whose adiabatic quantization yields a strictly positive rotor spacing, without introducing any Higgs field or explicit mass term. The unavoidable scaling $\Delta\sim(\text{dimensionless})/R$ is therefore separated from the nontrivial content: Gauss law (via $\Pi_T^{(D)}$), topology (meridian class), and the $\ZZ_3$ Berry twist enforce a nonzero spacing with a coefficient fixed by a Yang--Mills variational problem. Any Minkowski-space identification of $R$ requires additional physical input (e.g.\ a confinement length), which is stated explicitly.

\psec{Geometry, topology, and integer homology}{sec:geom}
Let $\Gamma\subset B_R\subset\RR^3$ be a smooth closed curve (a knot). Excise a thin open tubular neighborhood $\mathrm{Tub}(\Gamma)$; define the domain
\[
\cD := B_R\setminus \mathrm{Tub}(\Gamma).
\]
\textit{(i) Boundary description.} The boundary of the excised tube is a torus:
\[
\partial\!\mathrm{Tub}(\Gamma)\ \cong\ S^1\times S^1.
\]
We emphasize that $\partial\!\mathrm{Tub}(\Gamma)$---not $\Gamma$ itself---is the torus boundary component; the curve $\Gamma$ has empty boundary. Thus the full boundary is $\partial \cD = \partial B_R \;\cup\; \partial\!\mathrm{Tub}(\Gamma)$.

\textit{(ii) Meridian homology.} The abelianized loop class linking $\Gamma$ once is the meridian; one has
\[
H_1(\cD;\ZZ)\cong\ZZ.
\]
We will only use this integer homology (and the corresponding cohomology below); \emph{no identification of $\pi_1$ is required}.

\textit{(iii) Cohomology facts used later.} For a compact 3-manifold with boundary like $\cD$,
\[
H^1(\cD;\ZZ)\cong \ZZ,\qquad H^2(\cD;\ZZ)=0,
\]
(see App.~A via Alexander or Poincar\'e--Lefschetz duality). The vanishing $H^2$ ensures that closed 2-forms are exact; in the abelian seed we will use that there are no relative harmonic 1-forms, a point relevant in \Snameref{sec:holonomy} (cf.~\cite{TaylorPDE,EvansPDE} for elliptic/Hodge background).

\psec{Fields, function spaces, and boundary conditions}{sec:fields}
Let $A_\mu=A_\mu^aT^a$ be an SU(3) connection. The curvature and covariant derivative are
\begin{align}
F_{\mu\nu}&=\partial_\mu A_\nu-\partial_\nu A_\mu+\ii g[A_\mu,A_\nu],\label{eq:Fdef}\\
D_\mu &= \partial_\mu+\ii g[A_\mu,\cdot]\label{eq:Ddef}
\end{align}
with $\tr(T^aT^b)=\frac12\delta^{ab}$. Gauge transformations $h:\cD\to\mathrm{SU}(3)$ act by
\begin{align}
A_i&\mapsto h^{-1}A_i h+\tfrac{\ii}{g}\,h^{-1}\partial_i h,\qquad
D_i\mapsto h^{-1}D_i h.\label{eq:gauge}
\end{align}

\textit{Function spaces and BCs.} We take spatial one-forms $A=A_i\,\dd x^i\in W^{1,2}(\cD,\mathfrak{su}(3))$ and the scalar $A_0\in W^{1,2}_0(\cD,\mathfrak{su}(3))$ (Dirichlet on $\partial\cD$). For one-forms we impose \emph{relative} (electric-type) boundary conditions (covariantized):
\textit{Notation.} We reserve $H^k(\cD;\ZZ)$ for (co)homology groups, while Sobolev spaces are denoted $W^{k,2}(\cD)$ (so $W^{1,2}_0$ means Dirichlet trace).
\begin{equation}
\label{eq:relativeBC}
\iota_n A\big|_{\partial\cD}=0,\qquad 
\iota_n \, \star D A\big|_{\partial\cD}=0,
\end{equation}
and for scalars (Lie-algebra valued 0-forms) Dirichlet:
\begin{equation}
\phi\big|_{\partial\cD}=0,\qquad A_0\big|_{\partial\cD}=0.\label{eq:DirichletScalar}
\end{equation}
Here $\phi$ denotes a generic adjoint-valued scalar (e.g.\ the scalar potential in the covariant Helmholtz decomposition used below); we impose Dirichlet conditions on all such scalars, in particular on the Gauss-law potential $A_0$.
For PDE hygiene: traces are well-defined under $C^{1,1}$ boundaries with the regularity assumptions used below; alternatively one may work with smooth compactly supported fields and close the operators by density.

\textit{Adjoint identity and notation for backgrounds.} With the $L^2$ inner product $\ip{X}{Y}:=\int_{\cD}\tr(X_i Y_i)\,\dd^3x$ on one-forms, one has $D^\dagger=-D\!\cdot$ under \eqref{eq:relativeBC} and \eqref{eq:DirichletScalar} (derivation in App.~B). With Dirichlet scalars, $-D^2:=-(D\!\cdot D)$ on $W^{1,2}_0$ is positive, self-adjoint, and invertible (bounded domain, elliptic regularity \cite{TaylorPDE,EvansPDE}); denote
\[
G:=(D\!\cdot D)^{-1}:L^2(\cD)\to W^{2,2}(\cD)\cap W^{1,2}_0(\cD).
\]
When a static background $A^{(0)}$ is fixed, we write $D^{(0)}$ for its covariant derivative and $G^{(0)}$ for the corresponding Dirichlet inverse.

\psec{Covariant Dirichlet Helmholtz projector (idempotence, orthogonality, distance formula)}{sec:projector}
Define the projector on one-forms (here $G$ acts on Dirichlet scalars):
\begin{equation}
\Pi_T^{(D)}:= \mathbf{1} - D\,G\,D\!\cdot = \mathbf{1}+ D\,(-D^2)^{-1}D\!\cdot.\label{eq:PiT}
\end{equation}
\textit{Idempotence and divergence-free range} follow from $G(D\!\cdot D)=\mathbf{1}$, hence $(\Pi_T^{(D)})^2=\Pi_T^{(D)}$ and $D\!\cdot(\Pi_T^{(D)}Y)=0$. Orthogonality holds since for $Z=D\phi$ (Dirichlet $\phi$) and $X\in\ker D\!\cdot$,
$\ip{Z}{X}=\ip{D\phi}{X}=-\ip{\phi}{D\!\cdot X}=0$.
Thus $\Pi_T^{(D)}$ is the orthogonal projector onto $\ker D\!\cdot$, with $\|\Pi_T^{(D)}\|\le 1$ on $L^2$ and bounded $W^{1,2}\to W^{1,2}$.

\textit{Distance formula.} Every one-form decomposes $v=D\phi+\epsilon$ with $\phi\in W^{1,2}_0$ and $\epsilon\in\ker D\!\cdot$, uniquely with
\[
\epsilon=\Pi_T^{(D)}v,\quad
\norm{\epsilon}^2=\inf_{\phi'\in W^{1,2}_0}\norm{v-D\phi'}^2.
\]
(Uniqueness holds modulo covariantly harmonic relative 1-forms; see Assumption~A in \Snameref{sec:adiabatic}.)

\psec{Variational definition of the slow velocity (canonical inertia)}{sec:slow-variational}
Let $\alpha$ be the holonomy angle (\Snameref{sec:holonomy}). Define the \emph{slow velocity} at fixed $\alpha$ as the minimizer
\begin{equation}
\label{eq:var-problem}
X_\alpha \ =\ \arg\min_{X\in \mathcal V_\alpha}\ \frac{1}{2g^2}\,\norm{\Pi_T^{(D)} X}_{L^2(\cD)}^2,
\end{equation}
over the affine space $\mathcal V_\alpha$ of admissible velocities satisfying the \emph{holonomy-variation constraint}
\begin{equation}
\label{eq:hol-constraint}
\partial_\alpha U_\gamma\ =\ \ii\,(\AdV \, \cJ)\,U_\gamma,
\end{equation}
for any meridian $\gamma$ linking $\Gamma$ once (the corresponding \emph{linear functional} on $X$ is derived from the first-variation formula in App.~C). Equivalently—letting $U(s)$ be the parallel transporter along $\gamma$ at fixed $\alpha$—
\begin{equation}
\int_\gamma \tr\!\big((\AdV \, \cJ)\,U^{-1} X\,U\big)\ =\ \frac{1}{g}\,\tr(\cJ^2),\label{eq:scalar-constraint}
\end{equation}
which fixes the normalization of the constraint.

\textit{Uniqueness modulo covariant gradients.} The minimizer exists and is \emph{unique modulo} $\mathrm{Im}\,D$; i.e., unique in the quotient space $W^{1,2}(\cD)/\mathrm{Im}\,D$. Physically only $\Pi_T^{(D)}X$ matters. Indeed, for the seed obeying the constraint
\begin{equation}
\label{eq:seed}
X_\alpha^{\rm seed}\ :=\ \frac{1}{g}\,(\AdV \, \cJ)\,\omega,
\end{equation}
with a smooth closed 1-form $\omega$ supported away from the \emph{outer} boundary component $\partial B_R$ (but allowed to have support near the inner torus boundary $\partial\!\mathrm{Tub}(\Gamma)$) and normalized by $\oint_\gamma \omega=1$ (App.~C), minimization yields
\begin{equation}
\label{eq:minimizer}
\Pi_T^{(D)} X_\alpha\ =\ \Pi_T^{(D)} X_\alpha^{\rm seed}\,,
\end{equation}
so the physical electric field and the inertia are \emph{representative-independent}.

\psec{Gauss law, electric energy, and inertia}{sec:gauss}
Let $A_i^{(0)}$ be a smooth static background solving $D_\mu F^{\mu\nu}[A^{(0)}]=0$. Along a slow trajectory $\alpha(t)$ with velocity $X_\alpha$,
\begin{equation}
\partial_0 A_i \ =\ \dot\alpha\, (X_\alpha)_i.
\end{equation}
Gauss law $D_i F_{0i}=0$ with Dirichlet $A_0|_{\partial\cD}=0$ is solved by
\begin{equation}
D_i D_i A_0 = D_i \partial_0 A_i,\qquad A_0\big|_{\partial\cD}=0.\label{eq:A0Poisson}
\end{equation}
Hence, using $G=(D\!\cdot D)^{-1}$,
\begin{equation}
F_{0i}
=\Pi_T^{(D)}\partial_0 A_i.
\label{eq:F0iProj}
\end{equation}
The electric energy aligns as
\begin{align}
E_E
&=\frac{1}{2g^2}\int_{\cD}\tr(F_{0i}F_{0i})\,\dd^3x \notag\\
&=\frac{\dot\alpha^{\,2}}{2g^2}\,\ip{\Pi_T^{(D)} X_\alpha}{\Pi_T^{(D)} X_\alpha} \notag\\
&=\frac{\dot\alpha^{\,2}}{2}\,I_{\rm eff},\label{eq:Erot}
\end{align}

At quadratic adiabatic order the magnetic energy is stationary along the slow holonomy path and contributes only beyond the inertia definition; see Appendix~E.

with
\begin{align}
I_{\rm eff}
  &:= \frac{1}{g^2}\,\norm{\PT X_\alpha}^2 \notag\\
  &= \frac{1}{g^4}\,
     \ip{\PT\bigl((\AdV \, \cJ)\,\omega\bigr)}{\PT\bigl((\AdV \, \cJ)\,\omega\bigr)}.
     \label{eq:Ieff-seed}
\end{align}

\textit{A safe upper bound at quadratic adiabatic order.} Write $w:=(\AdV \,\cJ)\,\omega$ and take any smooth cut-off $\chi$ with $\chi|_{\partial\cD}=0$ and $(\nabla \chi)|_{\partial\cD}=0$. Define $v:=D^{(0)}(\chi \cJ)$. Using $\|\Pi_T^{(D)}\|\le 1$, the triangle inequality and the distance formula of \Snameref{sec:projector}, we obtain
\begin{align}
I_{\rm eff}
&= \frac{1}{g^4}\,\|\PT w\|_{L^2}^2 \notag\\
&\le \frac{1}{g^4}\Biggl[
\inf_{\phi\in W^{1,2}_0}
\Bigl\|D^{(0)}(\chi\cJ)-D^{(0)}\phi\Bigr\|_{L^2}^2
\notag\\
&\qquad\qquad
+ \Bigl\|w-D^{(0)}(\chi\cJ)\Bigr\|_{L^2}^2\Biggr] \nonumber
\\ &~~~~~~~~~~~~+ \mathcal O\!\bigl(\|A-A^{(0)}\|_{W^{1,2}}\bigr).
\label{eq:Ieff-distance}
\end{align}

The tightest upper bound is attained by solving the constrained infimum (Lagrange multiplier enforcing the meridian functional of App.~C). Since holonomy eigenvalues change along the slow motion (\Snameref{sec:holonomy}), the constrained seed $w=(\AdV\,\cJ)\,\omega$ (and hence $\Pi_T^{(D)}w$) cannot be $D$-exact; otherwise $\partial_\alpha U_\gamma=0$ by \eqref{eq:varU}. Hence $I_{\rm eff}>0$.

Interpretation of the non-$D$-exactness. 
The corrected statement is: it is the constrained seed/minimizer $w=(\AdV\,\cJ)\,\omega$ (not the auxiliary $v=D^{(0)}(\chi\cJ)$) that cannot be $D$-exact once the holonomy eigenvalues are required to vary. This is precisely why Gauss' law forces a nonzero transverse electric contribution and hence a strictly positive rotor inertia.
.

\psec{Holonomy control and the Berry twist}{sec:holonomy}
\textit{Closed 1-form.} There exists a smooth closed 1-form $\omega$ supported away from the outer boundary component $\partial B_R$ (but allowed to have support near the inner torus boundary $\partial\!\mathrm{Tub}(\Gamma)$) with
\begin{equation}
\int_\gamma \omega=1\quad \text{for any meridian }\gamma\ \text{linking }\Gamma \text{ once}.\label{eq:omega}
\end{equation}
Its construction uses $H^1(\cD)=\ZZ$ and $H^2(\cD)=0$ (App.~C).

\textit{Holonomy variation.} For a small variation $\delta A$, the path-ordered exponential obeys
\begin{equation}
\delta U_\gamma=\ii g\,U_\gamma\,\Big[\int_\gamma U^{-1}\,\delta A\,U\Big],\label{eq:varU}
\end{equation}
with $U$ the parallel transporter along the path (derivation in App.~C). Choose a thin interior ribbon supporting $\omega$ and fix \emph{axial gauge along that ribbon} so that the transporter is the identity there. With the seed \eqref{eq:seed}, $\oint_\gamma\omega=1$ gives
\begin{equation}
\partial_\alpha U_\gamma
=\ii\,(\AdV \, \cJ)\,U_\gamma,\qquad
U_\gamma(\alpha)=V e^{\ii\alpha \, \cJ}V^{-1}.\label{eq:Ualpha}
\end{equation}
All large gauge transformations we invoke are \emph{trivial on $\partial\cD$} and wrap the noncontractible meridian; thus they are compatible with the Dirichlet/relative boundary conditions. At the adiabatic level the Yang--Mills state is approximated by a Born--Oppenheimer form $|\Psi(t)\rangle\approx \psi(\alpha(t))\,|\Omega(\alpha(t))\rangle$, where $|\Omega(\alpha)\rangle$ is the instantaneous ground state of the fast transverse sector and $\psi$ is the effective one-dimensional rotor wavefunction for the gauge-invariant holonomy angle $\alpha$.Supporting material is organized as follows: Appendix~E (quadratic-order magnetic-energy stationarity), Appendix~F (bounds justifying $\tilde c_1=\mathcal O(1)$ in thin-tube geometries), and Appendix~G (conceptual clarification separating internal holonomy rotation from spacetime frame dragging).

Meaning of $\psi(\alpha)$. 
The appearance of the wavefunction is the collective-coordinate quantization statement: once $\alpha$ is identified as the slow, gauge-invariant holonomy modulus, the slow sector is a one-dimensional quantum mechanics on the circle $S^1_\alpha$, so physical slow states are represented by $\psi(\alpha)\in L^2(S^1_\alpha)$ (with a center-sector twist). In a fixed $\ZZ_3$ sector determined by the Wilson--’t Hooft algebra~\cite{Gaiotto2015,Kapustin2006}, they enforce the twisted boundary condition on slow wavefunctions:
\begin{equation}
\psi(\alpha+2\pi)=e^{2\pi \ii \, \nu/3}\,\psi(\alpha),\qquad \nu=0,1,2.\label{eq:twist}
\end{equation}
This is the \emph{Berry twist} for the 1D rotor and must not be confused with the 4D $\theta_{\rm YM}$.

\psec{Scaling and the commutator-dominated regime}{sec:scaling}
Set $\hbar=c=1$ in intermediate steps (restored in \Snameref{sec:benchmark}). Rescale $x=Ry$, $\widehat A_i(y):=R A_i(Ry)$. Then
\[
D_i=\frac{1}{R}\widehat D_i,\quad D\!\cdot=\frac{1}{R}\widehat D\!\cdot,\quad
\int_{\cD}\dd^3 x=R^3\int_{\widehat\cD}\dd^3y.
\]
A short computation gives $\Pi_T^{(D)}=\widehat\Pi_T^{(\widehat D)}$ because $G=(D\!\cdot D)^{-1}$ scales to $\widehat G=R^2 (\widehat D\!\cdot \widehat D)^{-1}$. Hence
\begin{equation}
I_{\rm eff}
=\frac{R}{g^4}\,
\ip{\widehat\Pi_T^{(\widehat D)}\,\widehat X_\alpha}{\widehat\Pi_T^{(\widehat D)}\,\widehat X_\alpha}_{\widehat\cD}
=: \tilde c_1\,\frac{R}{g^4},\qquad \tilde c_1=\mathcal O(1).\label{eq:scaling}
\end{equation}
For explicit geometric control (e.g.\ thin-tube regimes) we provide bounds showing $\tilde c_1$ remains $\mathcal O(1)$; see Appendix~F. Pointwise,
\[
\norm{D(\chi\cJ)}\ \le\ \norm{\partial(\chi\cJ)} + g\,\norm{[A,\chi\cJ]}.
\]
If the commutator term dominates define
\begin{equation}
\label{eq:epsilon}
\epsilon \;:=\; \frac{\norm{\partial(\chi\cJ)}_{L^2}}{g\,\norm{[A,\chi\cJ]}_{L^2}} \;\ll\; 1 .
\end{equation}
\noindent\emph{E.g. for backgrounds localized near the tube so $[A,\chi\cJ]$ lies where $\chi=1$.} Then $D(\chi\cJ)=\ii g [A,\chi\cJ]+\mathcal O(\epsilon g \, \norm{[A,\chi\cJ]})$ and
\begin{equation}
\label{eq:commdom}
I_{\rm eff}
= \frac{R}{g^2}\bigl(c_1+\mathcal O(\epsilon)\bigr),
\qquad
c_1 := \bigl\|\widehat\Pi_T^{(\widehat D)}[\widehat A,\chi\cJ]\bigr\|_{\widehat\cD}^{\!2} > 0.
\end{equation}

\psec{Projector stability, vector gap, and adiabatic decoupling}{sec:adiabatic}
Let $H_{\cD}$ be the YM Hamiltonian in the fixed sector. Define the \emph{transverse vector gap}
\begin{equation}
\Lambda_T:=\lambda_1\!\left(-D^2\ \text{on}\ \ker D\!\cdot\ \text{with \eqref{eq:relativeBC}}\right)>0.\label{eq:LambdaT}
\end{equation}
\textbf{Assumption A (transverse irreducibility).} There is no nontrivial $X$ with $DX=0$ and $D\!\cdot X=0$ obeying the relative boundary conditions~\eqref{eq:relativeBC}. Under Assumption A one has $\Lambda_T>0$ by integration by parts and ellipticity (App.~D; cf.~\cite{TaylorPDE,EvansPDE}). By elliptic scaling, $\Lambda_T(\cD_R)=\Lambda_T(\cD_1)/R^2$, hence $\sqrt{\Lambda_T}\sim 1/R$.

\textit{Projector stability (operator-norm estimate).} Assume $\partial\cD$ is $C^{1,1}$ and $A,A^{(0)}\in W^{1,\infty}(\cD)$ with
\[
\|A\|_{W^{1,\infty}}+\|A^{(0)}\|_{W^{1,\infty}}\ \le\ M.
\]
Here $M$ is an \emph{a priori} bound fixing a bounded subset of backgrounds; once $M$ is fixed, the constant $C(M,\cD)$ is finite and depends only on $(M,\cD)$ (and is scale-invariant after rescaling $\cD_R\mapsto\cD_1$).
Then
\begin{equation}
\begin{aligned}
\big\|\Pi_T^{(D)}-\Pi_T^{(D^{(0)})}\big\|_{L^2\to L^2}
\ &\le\ C(M,\cD)\,
\\
&\quad\times \|A-A^{(0)}\|_{W^{1,2}(\cD)} .
\end{aligned}
\label{eq:projstab}
\end{equation}

Justification of the $C(M,\cD)$-dependence and the norm control. 
With $L_A:=D_A\!\cdot D_A$ on $W^{1,2}_0(\cD)$ (Dirichlet) and $G_A:=L_A^{-1}$, the resolvent identity
$G_A-G_{A^{(0)}}=G_A\,(L_{A^{(0)}}-L_A)\,G_{A^{(0)}}$
combined with elliptic bounds on $\cD$ yields a Lipschitz estimate in $\|A-A^{(0)}\|_{W^{1,2}}$ on bounded $W^{1,\infty}$ sets (encoded by $M$). Substituting into $\Pi_T^{(D)}=\mathbf 1-D_AG_A D_A\!\cdot$ gives Eq.~\eqref{eq:projstab}.

\textit{Adiabatic control and Rayleigh bound.} For slowly varying $\alpha(t)$, a Kato/Feshbach--Schur reduction \cite{Simon1983} yields leakage into fast transverse modes bounded by
More precisely, if $|\Psi(t)\rangle$ is the exact state and $|\Omega(\alpha(t))\rangle$ the instantaneous ground state of the fast sector, we define the leakage probability $\varepsilon_{\rm ad}(t):=\|\Pi_\perp(\alpha(t))|\Psi(t)\rangle\|^2$ with $\Pi_\perp(\alpha):=\mathbf 1-|\Omega(\alpha)\rangle\langle\Omega(\alpha)|$.
\begin{equation}
\varepsilon_{\rm ad}\ \le\ C\,\Big(\frac{\dot\alpha}{\sqrt{\Lambda_T}}\Big)^{\!2},\label{eq:epsad}
\end{equation}
for a domain-dependent $C=\mathcal O(1)$, provided the self-adjointness domain of $H_{\cD}$ is fixed by \eqref{eq:relativeBC}. Let $|\Omega(\alpha)\rangle$ be the instantaneous ground state and set $|\psi_1\rangle:=\Pi_\perp\,\partial_\alpha|\Omega\rangle|_{\alpha=0}$. Then
\begin{equation}
\label{eq:variational}
\begin{aligned}
\frac{\langle \psi_1|H_{\cD}|\psi_1\rangle}{\langle \psi_1|\psi_1\rangle}
  &= \Delta_{\rm rotor}\bigl(1+\mathcal{O}(\varepsilon_{\rm ad})\bigr),\\
\lambda_{\min}^{(>0)}(H_{\cD})
  &\le \Delta_{\rm rotor}\bigl(1+\mathcal{O}(\varepsilon_{\rm ad})\bigr).
\end{aligned}
\end{equation}
Moreover, along an adiabatic trajectory $A(t)=A(\alpha(t))$ one has $\partial_0A=(\partial_\alpha A)\,\dot\alpha$, and projector stability \eqref{eq:projstab} controls the difference $\Pi_T^{(D(\alpha))}-\Pi_T^{(D^{(0)})}$ by $\|A(\alpha)-A^{(0)}\|_{W^{1,2}}$. On a time window where $|\alpha|$ remains small (with $A^{(0)}:=A(\alpha=0)$), this error enters only beyond quadratic adiabatic order, so evaluating $I_{\rm eff}$ with $D^{(0)}$ is consistent at the order used here.

\psec{Rotor quantization with $\ZZ_3$ Berry twist}{sec:rotor}
On the slow manifold, the effective mechanics is
\begin{equation}
L=\tfrac12 I_{\rm eff}\dot\alpha^2+\frac{\theta}{2\pi}\dot\alpha,\qquad
p=\frac{\partial L}{\partial \dot\alpha}=I_{\rm eff}\dot\alpha+\frac{\theta}{2\pi}.
\end{equation}
The $\ZZ_3$ twist \eqref{eq:twist} implies $\theta=2\pi\delta$ with $\delta\in\{0,1/3,2/3\}$. Quantization gives $p\in \ZZ-\delta$, hence energies
\begin{equation}
H=\frac{1}{2I_{\rm eff}}\,(n-\delta)^2,\qquad
\Delta=
\begin{cases}
\dfrac{1}{2I_{\rm eff}},&\delta=0,\\[4pt]
\dfrac{1}{6I_{\rm eff}},&\delta=\tfrac13,\tfrac23 .
\end{cases}\label{eq:gap-values}
\end{equation}
This $\theta$ is the \emph{Berry twist} of the rotor, not the 4D $\theta_{\rm YM}$.

\psec{Quantitative benchmark and two-angle SU(3) generalization}{sec:benchmark}
Restoring units, for $\delta=\tfrac13,\tfrac23$,
\begin{equation}
\Delta=\frac{1}{6I_{\rm eff}}
=\frac{g^4}{6\tilde c_1}\,\frac{\hbar c}{R}.\label{eq:bench}
\end{equation}
This benchmark is an \emph{illustrative calibration}: choosing $R$ at a confinement-scale length produces a hadronic-scale upper bound, but $R$ itself is not fixed by the present finite-domain analysis. Taking $R=1\,\mathrm{fm}$ and $\alpha_s(\kappa/R)=0.4$ ($g^2=4\pi\alpha_s\simeq 5.0266$), $g^4/6\simeq 4.21$ and $\hbar c/R\simeq 197.327\,\mathrm{MeV}$, hence
\begin{equation}
\Delta_{\rm var}\ \approx\ \frac{830}{\tilde c_1}\ \mathrm{MeV},\label{eq:830}
\end{equation}
with $\tilde c_1=\mathcal O(1)$. A moderate scale variation $\kappa\in[0.5,2]$ keeps \eqref{eq:830} stable at the $\sim 20\%$ level.
For two Cartan angles $H_a$ ($\tr H_aH_b=\tfrac12\delta_{ab}$), define
\begin{equation}
I_{ab}=\frac{1}{g^4}\,\ip{\Pi_T^{(D^{(0)})}D^{(0)}(\chi H_a)}{\Pi_T^{(D^{(0)})}D^{(0)}(\chi H_b)}.
\end{equation}
Along unit $\hat u$, $\Delta(\hat u)=1/\big(2\,\hat u^{\!\top}I\,\hat u\big)\ge 1/\big(2\,\lambda_{\max}(I)\big)$. The $\ZZ_3$ center fixes a \emph{single} twist on the Cartan torus; the pair $(\delta_1,\delta_2)$ therefore lies on one center-induced shift class (not freely independent componentwise). The tightest variational upper bound uses the principal axis of $I$.

\psec{Relation to bags, soliton rotors, and broader context}{sec:relation}
Bag/cavity spectra~\cite{Chodos1974} quantize finite-volume modes but without a Berry twist. Soliton rotors (e.g.\ Skyrmions~\cite{AdkinsNappiWitten1983}) rotate localized lumps. Here the rotor is a \emph{gauge-invariant holonomy mode} supported by topology and Gauss law, with a \emph{$\ZZ_3$ Berry} shift fixed by the center sector. The construction is consistent with generalized-symmetry pictures of confinement~\cite{Gaiotto2015,Kapustin2006}. A short conceptual remark on why the internal holonomy motion does not imply spacetime rotation (frame dragging) is given in Appendix~G.

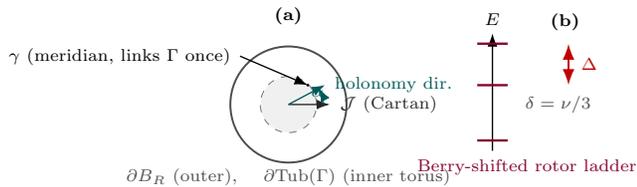
\begin{figure}[t]
\centering
\resizebox{\columnwidth}{!}{%
\begin{tikzpicture}[x=1cm,y=1cm,>=Latex,
  every node/.style={font=\scriptsize},
  axis/.style={-Latex, line width=0.55pt},
  ring/.style={line width=0.8pt},
  ringd/.style={line width=0.8pt, dashed},
  level/.style={line width=0.95pt},
  gap/.style={<->, line width=0.8pt},
  lab/.style={font=\scriptsize},
  panel/.style={font=\bfseries\scriptsize}
]

\begin{scope}[shift={(-2.65,0.62)}]
  \node[panel] at (0,1.28) {(a)};

  \draw[ring, draw=black!70] (0,0) circle (0.84);
  \draw[ringd, draw=black!55] (0,0) circle (0.40);

  \fill[black!6] (0,0) circle (0.40);

  \fill[black] (0.28,0.28) circle (0.020);
  \draw[axis, draw=black] (-0.70,0.70) -- (0.28,0.28);
  \node[lab, anchor=east, text=black] at (-0.72,0.70)
    {$\gamma$ (meridian, links $\Gamma$ once)};

  \draw[axis, draw=black!85] (0,0)--(0.60,0);
  \node[lab, anchor=west, text=black!85] at (0.63,0.00) {$\cJ$ (Cartan)};

  \begin{scope}[rotate=28]
    \draw[axis, draw=teal!70!black] (0,0)--(0.60,0);
    \node[lab, anchor=west, text=teal!70!black] at (0.63,0.00) {holonomy dir.};
  \end{scope}
  \draw[axis, draw=teal!70!black] (0.50,0) arc[start angle=0, end angle=28, radius=0.50];
  \node[lab, text=teal!70!black] at (0.42,0.17) {$\alpha$};

  \node[lab, text=black!70] at (0,-1.05)
    {$\partial B_R$ (outer), \quad $\partial\mathrm{Tub}(\Gamma)$ (inner torus)};
\end{scope}

\begin{scope}[shift={(1.35,0.60)}]
  \node[panel] at (0,1.22) {(b)};

  \draw[axis] (-1.05,-0.65)--(-1.05,1.05);
  \node[lab, anchor=south] at (-1.05,1.05) {$E$};

  \draw[level, draw=purple!70!black] (-1.27,-0.50)--(-0.83,-0.50);
  \draw[level, draw=purple!70!black] (-1.27,0.30)--(-0.83,0.30);
  \draw[level, draw=purple!70!black] (-1.27,0.90)--(-0.83,0.90);

  \draw[gap, draw=red!75!black] (0.05,0.30)--(0.05,0.90);
  \node[lab, text=red!75!black, anchor=west] at (0.12,0.60) {$\Delta$};

  \node[lab, text=purple!70!black] at (-0.55,-0.88) {Berry-shifted rotor ladder};

  \node[lab, text=black!70] at (-0.10,0.02) {$\delta=\nu/3$};
\end{scope}

\end{tikzpicture}
}%
\caption{(a) Punctured domain $\cD=B_R\setminus\mathrm{Tub}(\Gamma)$: the inner dashed circle schematizes the excised tube boundary, and $\gamma$ is a meridian linking $\Gamma$ once. The holonomy angle $\alpha$ parametrizes the gauge-invariant modulus in a fixed $\ZZ_3$ sector. (b) Berry-shifted rotor spectrum with level spacing $\Delta$.}
\label{fig:schematic}
\end{figure}

\paragraph*{Conclusions.}
A single, gauge-invariant holonomy mode on a punctured SU(3) domain yields a Berry-shifted rotor with strictly positive spacing. Defining the slow velocity variationally and enforcing Gauss law via a covariant Dirichlet Helmholtz projector fixes a canonical inertia and produces a \emph{finite-domain} variational \emph{upper} bound on the YM spectrum, with a dimensionless adiabatic error controlled by the \emph{transverse vector gap}. This gives a concrete, local ``mass without mass'' mechanism. Next steps: sharper analytic bounds for $\tilde c_1,c_1$ in canonical geometries and lattice tests on punctured domains.

\section*{Acknowledgments}

The author gratefully acknowledges Nader Inan, Jos\'e Rodal, and Douglas Singleton for valuable email discussions on the concept of \emph{mass without mass}.

\appendix

\section*{Appendix A: Cohomology and duality statements}
For $\cD=B_R\setminus \mathrm{Tub}(\Gamma)$ (knot complement in a ball), Alexander duality or Poincar\'e--Lefschetz duality gives $H^1(\cD;\ZZ)\cong \ZZ$ (generated by the meridian class) and $H^2(\cD;\ZZ)=0$. Concretely, a meridian loop linking $\Gamma$ once is nontrivial, while any closed 2-form on $\cD$ is exact.

\section*{Appendix B: Integration by parts and $D^\dagger=-D\!\cdot$}
Let $X$ be a scalar, $Y$ a one-form. Write
\[
\tr(D_iX\,Y_i)=\partial_i\tr(XY_i)-\tr\left(X\,\partial_i Y_i\right)+\ii g\,\tr\left([A_i,X]Y_i\right).
\]
Integrate on $\cD$ and apply the divergence theorem; the boundary term is $\int_{\partial\cD}\tr(X\,\iota_n Y)\,\dd S$, which vanishes by $\iota_n Y|=0$ and $X|=0$. Finally, $\tr([A_i,X]Y_i)=\tr\left(X[Y_i,A_i]\right)$ and $D\!\cdot Y=\partial_iY_i+\ii g[Y_i,A_i]$.

\section*{Appendix C: Holonomy variation: first-variation formula and closed form $\omega$}
Let $\gamma$ be a parameterized loop, $U(s)$ the transporter along $\gamma|_{[0,s]}$. Then
\[
\frac{\dd}{\dd s} U(s)=\ii g\,\dot\gamma^i(s)\,A_i(\gamma(s))\,U(s).
\]
For a small variation $\delta A$, Duhamel expansion yields
\[
\delta U(1)=\ii g\!\int_0^1 U(1)U^{-1}(s)\,\dot\gamma^i(s)\,\delta A_i(\gamma(s))\,U(s)\,\dd s,
\]
equivalently \eqref{eq:varU}. To construct $\omega$, start from the angular form $\omega_0$ in a tube around $\gamma$ with $\oint_\gamma\omega_0=1$, multiply by a bump $b$ supported away from the \emph{outer} boundary $\partial B_R$ (but not required to vanish near $\partial\!\mathrm{Tub}(\Gamma)$), and then correct the failure of closedness $d(b\omega_0)=db\wedge\omega_0$ by introducing a smooth one-form $\eta$ satisfying $d\eta=-db\wedge\omega_0$ (possible because $H^2(\cD)=0$). Set  $\omega=b\omega_0+\eta$.
Because $\gamma$ links the excised tube, any spanning surface for $\gamma$ necessarily meets $\partial\!\mathrm{Tub}(\Gamma)$; thus Stokes' theorem does not force $\oint_\gamma\omega=0$ when $\omega$ vanishes only near the outer boundary $\partial B_R$.

\section*{Appendix D: Projector stability and positivity of $\Lambda_T$}
Assume $A,A^{(0)}\in W^{1,\infty}$ and $\partial\cD$ is $C^{1,1}$. Then the Dirichlet inverses $G=(D\!\cdot D)^{-1}$ and $G^{(0)}=(D^{(0)}\!\cdot D^{(0)})^{-1}$ are bounded $L^2\to H^2\cap W^{1,2}_0$ with bounds depending on $\|A\|_{W^{1,\infty}}$, $\|A^{(0)}\|_{W^{1,\infty}}$ and $\cD$ \cite{TaylorPDE,EvansPDE}. Writing
\[
\Pi_T^{(D)}-\Pi_T^{(D^{(0)})}= -DGD\!\cdot + D^{(0)}G^{(0)}D^{(0)}\!\cdot,
\]
insert $D=D^{(0)}+\ii g[A-\!A^{(0)},\cdot]$ and $G=G^{(0)}+G(D\!\cdot D- D^{(0)}\!\cdot D^{(0)})G^{(0)}$; bound each term in $L^2\to L^2$ norm by submultiplicativity and the $L^\infty$ control on $A$. For $\Lambda_T>0$, if $-D^2 X=0$ and $D\!\cdot X=0$, integrate $\ip{X}{-D^2 X}=\norm{DX}^2$ and use \eqref{eq:relativeBC}; hence $DX=0$ and Assumption~A forces $X=0$.

\section*{Appendix E: Magnetic energy at quadratic order}
Let $E_B=\frac{1}{2g^2}\int\tr(F_{ij}F^{ij})$. Along the slow path, $\frac{\dd}{\dd \alpha}E_B|_{\alpha=0}=0$ using $D_iF^{ij}[A^{(0)}]=0$, Bianchi identity, relative BCs \eqref{eq:relativeBC}, and the collar condition for $\chi$; hence $E_B=\mathcal O(\dot\alpha^2)$ and does not affect $I_{\rm eff}$ at quadratic adiabatic order.

\section*{Appendix F: Bounds for $\tilde c_1$ on thin tubes}
On $\widehat\cD=B_1\setminus \widehat\Gamma$ with a thin tube $\widehat\Gamma$ of radius $\rho\ll 1$, take a smooth center-vortex-like $A^{(0)}$ supported near $\widehat\Gamma$ and $\chi$ equal to $1$ on a linking torus, vanishing near $\partial\widehat\cD$. Poincar\'e/trace inequalities plus $\norm{\Pi_T}\le 1$ imply $\kappa_{\rm low}(\rho)\le \tilde c_1\le \kappa_{\rm up}(\rho)$ with $\kappa_{\rm low},\kappa_{\rm up}=\mathcal O(1)$ as $\rho\to 0^+$, justifying \eqref{eq:scaling}.

\section*{Appendix G: Frame dragging and G\"odel contrast (conceptual)}
The slow coordinate $\alpha(t)$ rotates an \emph{internal} SU(3) holonomy on flat spacetime; it does not introduce spacetime rotation. Any gravitomagnetic (frame-dragging) effect would have to come from the Yang--Mills stress tensor via general relativity. The relevant source is the total momentum
\[
P_i=\int_{\cD} T_{0i}\,\dd^3x,\qquad T_{0i}=\frac{1}{g^2}\,\tr(\bm E\times \bm B)_i.
\]
With $n^iF_{i0}|_{\partial B_R}=0$, Dirichlet $A_0$, Gauss-law projection for $F_{0i}$, and a static background satisfying $D_\mu F^{\mu i}[A^{(0)}]=0$, the $\mathcal O(\dot\alpha)$ contribution cancels, so
\[
P_i=0+\mathcal O(\dot\alpha^2).
\]
Physically, the slow holonomy motion generates localized color-electric patterns that circulate but carry no net linear momentum through the boundary; hence no macroscopic frame dragging beyond ordinary sources (cf.~\cite{GPB2011,Ciufolini:2016ntr}). This contrasts with G\"odel’s spacetime, where inertial frames rotate due to the geometry itself \cite{Godel1949}.

\begin{thebibliography}{99}

\bibitem{Wilczek1999}
F.~Wilczek,
Mass without mass I: Most of matter,
Phys.~Today \textbf{52}(11), 11--13 (1999),
\url{https://doi.org/10.1063/1.882879}.

\bibitem{Wilczek2000}
F.~Wilczek,
Mass without mass II: The medium is the mass-age,
Phys.~Today \textbf{53}(1), 13--14 (2000),
\url{https://doi.org/10.1063/1.882927}.

\bibitem{Wilson1974}
K.~G.~Wilson,
Confinement of quarks,
Phys.~Rev.~D \textbf{10}, 2445--2459 (1974),
\url{https://doi.org/10.1103/PhysRevD.10.2445}.

\bibitem{Greensite2011}
J.~Greensite,
\textit{An Introduction to the Confinement Problem}
(Springer, 2011),
\url{https://doi.org/10.1007/978-3-642-14382-3}.

\bibitem{thooft1978}
G.~'t Hooft,
On the phase transition towards permanent quark confinement,
Nucl.~Phys.~B \textbf{138}, 1--25 (1978),
\url{https://doi.org/10.1016/0550-3213(78)90153-0}.

\bibitem{Gaiotto2015}
D.~Gaiotto, A.~Kapustin, N.~Seiberg and B.~Willett,
Generalized Global Symmetries,
JHEP \textbf{02}, 172 (2015),
\url{https://doi.org/10.1007/JHEP02(2015)172}.

\bibitem{Kapustin2006}
A.~Kapustin,
Wilson--'t Hooft operators in four-dimensional gauge theories and S-duality,
Phys.~Rev.~D \textbf{74}, 025005 (2006),
\url{https://doi.org/10.1103/PhysRevD.74.025005}.

\bibitem{AliEPL2025}
M.~Ali and A.~F.~Ali,
Deriving the cosmological constant and nature's constants from SU(3) confinement volume,
EPL (Europhys.~Lett.) \textbf{151}, 39002 (2025),
\url{https://doi.org/10.1209/0295-5075/adf509}.

\bibitem{AliSymmetry2025}
A.~F.~Ali,
Unbreakable SU(3) atoms of vacuum energy: A solution to the cosmological constant problem,
Symmetry \textbf{17}(6), 888 (2025),
\url{https://doi.org/10.3390/sym17060888}.

\bibitem{YM}
C.~N.~Yang and R.~L.~Mills,
Conservation of isotopic spin and isotopic gauge invariance,
Phys.~Rev.~\textbf{96}, 191--195 (1954),
\url{https://doi.org/10.1103/PhysRev.96.191}.

\bibitem{TaylorPDE}
M.~E.~Taylor,
\textit{Partial Differential Equations II: Qualitative Studies of Linear Equations}, 2nd ed.
(Springer, 2011),
\url{https://doi.org/10.1007/978-1-4419-7052-7}.

\bibitem{EvansPDE}
L.~C.~Evans,
\textit{Partial Differential Equations}, 2nd ed.
(American Mathematical Society, 2010),
\url{https://doi.org/10.1090/gsm/019}.

\bibitem{Chodos1974}
A.~Chodos, R.~L.~Jaffe, K.~Johnson, C.~B.~Thorn and V.~F.~Weisskopf,
New extended model of hadrons,
Phys.~Rev.~D \textbf{9}, 3471--3495 (1974),
\url{https://doi.org/10.1103/PhysRevD.9.3471}.

\bibitem{AdkinsNappiWitten1983}
G.~S.~Adkins, C.~R.~Nappi and E.~Witten,
Static properties of nucleons in the Skyrme model,
Nucl.~Phys.~B \textbf{228}, 552--566 (1983),
\url{https://doi.org/10.1016/0550-3213(83)90559-X}.



\bibitem{Simon1983}
B.~Simon,
Holonomy, the quantum adiabatic theorem, and Berry's phase,
Phys.~Rev.~Lett.~\textbf{51}, 2167--2170 (1983),
\url{https://doi.org/10.1103/PhysRevLett.51.2167}.

\bibitem{GPB2011}
C.~W.~F.~Everitt \textit{et al.},
Gravity Probe B: Final Results of a Space Experiment to Test General Relativity,
Phys.~Rev.~Lett.~\textbf{106}, 221101 (2011),
\url{https://doi.org/10.1103/PhysRevLett.106.221101}.

\bibitem{Ciufolini:2016ntr}
I.~Ciufolini \textit{et al.},
A test of general relativity using the LARES and LAGEOS satellites and a GRACE Earth gravity model,
Eur.~Phys.~J.~C \textbf{76}, 120 (2016),
\url{https://doi.org/10.1140/epjc/s10052-016-3961-8}.

\bibitem{Godel1949}
K.~G\"{o}del,
An example of a new type of cosmological solution of Einstein's field equations of gravitation,
Rev.~Mod.~Phys.~\textbf{21}, 447--450 (1949),
\url{https://doi.org/10.1103/RevModPhys.21.447}.
\end{thebibliography}
\end{document}